\documentclass[conference]{IEEEtran}

\usepackage[T1]{fontenc}
\usepackage[utf8]{inputenc}

\usepackage{amsmath}
\usepackage{amssymb}
\usepackage{amsfonts}
\usepackage{bm}                  

\usepackage{graphicx}
\graphicspath{{Figures/}}        
\usepackage[dvipsnames]{xcolor}

\usepackage{booktabs}            
\usepackage{multirow}
\usepackage{tabularx}

\usepackage{algorithm}
\usepackage{algpseudocode}

\usepackage{cite}                
\usepackage[colorlinks=true,
            linkcolor=blue,
            citecolor=blue,
            urlcolor=blue]{hyperref}

\usepackage{balance}             
\usepackage{microtype}           
\usepackage{url}
\usepackage{booktabs}

\usepackage{xcolor}

\title{Atlas: Optimizing Deployment of Compound AI Workflows on Heterogeneous Clusters}

\author{
  \IEEEauthorblockN{Milos Gravara}
  \IEEEauthorblockA{
    Distributed Systems Group\\
    TU Wien\\
    m.gravara@dsg.tuwien.ac.at
  }\and
  \IEEEauthorblockN{Andrija Stanisic}
  \IEEEauthorblockA{
    Distributed Systems Group\\
    TU Wien\\
    a.stanisic@dsg.tuwien.ac.at
  }
  \and
  \IEEEauthorblockN{Stefan Nastic}
  \IEEEauthorblockA{
    Distributed Systems Group\\
    TU Wien\\
    s.nastic@dsg.tuwien.ac.at
  }
}

\begin{document}

\maketitle

\begin{abstract}
  Compound AI workflows are increasingly used to serve complex AI tasks by coordinating multiple AI models and software components. This approach enables deployment flexibility, as each workflow stage can expose different model variants and resource requirements, but it also expands the deployment choices. A deployment must choose an execution plan that selects AI models for each compound AI workflow stage and places them on a heterogeneous cluster in order to satisfy SLOs. Deployment optimizers therefore need estimates to compare many candidate plans and identify feasible ones.  System metrics can often be profiled per stage and composed according to workflow topology, but accuracy cannot, as errors and information loss at upstream stages affect the accuracy of downstream stages. Existing approaches either profile complete configurations end to end, which scales poorly, or use product-based accuracy surrogates that treat stages as independent and can misrank candidate plans. We introduce Atlas, a framework for optimizing compound AI deployments under SLO constraints. Atlas uses MAP, a Markovian Accuracy Predictor, to estimate configuration accuracy from local conditional accuracy transitions between adjacent workflow stages. MAP discretizes intermediate outputs into accuracy buckets and composes transition profiles according to workflow topology, giving the optimizer an accuracy estimate without exhaustive end-to-end profiling. Atlas formulates execution-plan selection as a mixed-integer linear program that maximizes predicted accuracy subject to SLOs. Across four compound AI workflows, MAP achieves Spearman correlation up to 0.947 while reducing profiling cost by up to 2.6$\times$ relative to exhaustive end-to-end profiling. Guided by MAP, the Atlas optimizer selects execution plans within 0.03 of oracle accuracy while reducing deployment cost by up to 42\% through heterogeneous placement.
\end{abstract}

\begin{IEEEkeywords}
Compound AI, Model Selection, Deployment Optimization, Distributed Inference
\end{IEEEkeywords}

\section{Introduction}

The field of Artificial Intelligence (AI) is shifting from deploying monolithic AI models towards Compound AI systems. Compound AI represents a distributed intelligence approach combining multiple specialized AI models with software components into workflows, where each stage represents a single model or component invocation, orchestrated to solve various AI tasks~\cite{compound-ai-blog, Gravara2025ANC, Gravara2026Compass, Khattab2023DSPy, Wu2023AutoGen, Yao2022ReAct}. This approach offers practical advantages for reliability, scalability, and efficiency, enabling control over model outputs, component-specific adjustments, and adaptation to changing conditions~\cite{chen2023frugalgptuselargelanguage, HybridLLM, TOPERA, 10.1145/3770501.3770531, gravara2026plaiground, gravara2026design}.

These advantages are not obtained by the workflow structure alone. A Compound AI workflow must be instantiated as an execution plan before deployment~\cite{IPA, Ahmad2024Loki, Gravara2026Compass}. Such a plan selects the model variant, runtime parameters and hardware placement of each workflow stage to satisfy given Service Level Objectives (SLOs), which typically include latency, throughput, and cost constraints~\cite{IPA, Ahmad2024Loki, Jiang2018Chameleon, Jellyfish}. As production workflows grow in the number of stages, these choices create a combinatorial space of execution plans~\cite{optimas, chen2025llmselector}. Each plan occupies a different point in the accuracy-performance-cost space, which often compete~\cite{Zhang2020ModelSwitching, Mendoza2024RAMSIS, Romero2021INFaaS, Proteus, Wu2022JellyBean}. This means that, for example, selecting larger AI models per stage may improve overall accuracy, but typically increases latency and cost. On the other hand, plans including cheaper or faster AI models may satisfy SLOs, but often at the expense of workflow accuracy. Therefore, deployment optimization amounts to jointly selecting model variants per stage and mapping them onto heterogeneous hardware, managing the resulting trade-offs to find plans that maximize accuracy among those feasible under the given SLOs.

A common formulation of this optimization problem requires estimating accuracy and system behavior for each candidate execution plan before selection~\cite{IPA, Ahmad2024Loki, Wu2022JellyBean, Crankshaw2020InferLine}. Latency, throughput, and cost are mainly tractable in this setting because they can often be profiled for individual stage variants on target hardware and composed according to the workflow topology~\cite{Ahmad2024Loki, Proteus, IPA, Crankshaw2020InferLine}. Yet, such process cannot be applied to estimate workflow accuracy. As output of one stage becomes the input to downstream stages, errors and information loss introduced upstream can change the accuracy distribution of later stages~\cite{optimas, chen2025llmselector, Gravara2026Compass}. Workflow-level accuracy therefore depends on how quality propagates through the execution plan, inducing a challenge in estimating plan accuracy before deployment. 

Existing approaches commonly estimate workflow-level accuracy in one of two ways. End-to-end profiling evaluates complete execution plans directly and provides faithful measurements for plan selection~\cite{Ahmad2024Loki, Gravara2026Compass, Proteus, Wu2022JellyBean}. This captures interactions between stages, but each added variant, parameter, or stage requires additional complete workflow evaluations, making exhaustive profiling intractable as workflows grow in depth and variant count. To reduce profiling cost, recent work has constructed surrogate accuracy models by composing per-stage accuracy estimates, often as products of individual stage accuracies~\cite{IPA, chen2025llmselector}. Such surrogate models are efficient, but they treat stage contributions as largely independent and therefore do not capture how upstream errors or information loss affect downstream behavior. As a result, existing methods either preserve interaction fidelity at high profiling cost or reduce cost through assumptions that weaken accuracy estimation across multi-stage workflows.

In this paper, we introduce Atlas, a framework that optimizes model selection and hardware mapping for Compound AI workflows under SLO constraints. Atlas selects execution plans that maximize predicted task accuracy while satisfying latency, throughput, memory, and cost requirements. To estimate workflow accuracy without exhaustive end-to-end profiling, Atlas profiles how model choices at one workflow stage affect the output quality of the next. The Markovian Accuracy Predictor (MAP) then discretizes these quality signals into buckets and composes the resulting transition profiles according to the workflow topology, predicting end-to-end accuracy for any candidate configuration. The Atlas optimizer combines these accuracy predictions with system performance profiles in a mixed-integer linear program (MILP), producing an execution plan that specifies model selection and hardware mapping for each workflow stage.


The main contributions of this work are:

\begin{itemize}
  \item \textbf{Atlas} - A novel framework for deployment optimization of Compound AI workflows on heterogeneous clusters. Atlas takes a workflow specification, candidate model variants, a calibration dataset, and SLOs as input, and produces an execution plan specifying model variant selection and hardware placement per workflow stage. It separates accuracy estimation from system profiling, keeping both tractable, and supports linear, routed, loop, and composed workflow topologies.

  \item \textbf{MAP} - A \textbf{M}arkovian \textbf{A}ccuracy \textbf{P}redictor that estimates configuration accuracy from local conditional quality transitions between adjacent workflow stages. MAP discretizes intermediate outputs into quality buckets and composes transition profiles according to workflow topology using three operators: linear pipelines, routed workflows, and feedback loops. Across four evaluated workflows, MAP achieves the strongest ranking correlation among evaluated predictors, with Spearman $\rho$ = 0.947 on RAG, $\rho$ = 0.921 on RAG with routing, $\rho$ = 0.783 on RAG with self-refinement, and $\rho$ = 0.882 on the full composed workflow, while reducing profiling cost by 2.6$\times$ on our largest measured workflow, growing to over 80$\times$ in synthetic projection.

  \item \textbf{Atlas Plan Optimizer} - An execution plan optimizer that formulates plan selection as a Mixed-Integer Linear Program (MILP) over model variant selection and hardware placement, jointly optimizing predicted accuracy subject to latency, throughput, memory, and cost constraints. The predicted accuracy objective is supplied by MAP and evaluated over all candidate configurations before the MILP is invoked, keeping the objective linear. On a homogeneous cluster, Atlas selects execution plans within 0.03 of oracle accuracy across all evaluated loads and SLOs, while baseline approaches fall by up to 0.45. On a heterogeneous cluster, Atlas matches oracle accuracy at up to 42\% lower deployment cost than single-tier strategies.
\end{itemize}

The remainder of this paper is organized as follows. Section~\ref{sec:motivation} motivates the accuracy-estimation problem. Section~\ref{sec:framework-overview} presents the Atlas framework and system model. Section~\ref{sec:accuracy-model} introduces MAP. Section~\ref{sec:optimizer} formulates the MILP. Section~\ref{sec:evaluation} evaluates Atlas, Section~\ref{sec:related} reviews related work, Section~\ref{sec:limitations} discusses design choices and Section~\ref{sec:conclusion} concludes.

\section{Motivation}
\label{sec:motivation}


This section motivates the accuracy-estimation problem in Compound AI deployment optimization. We first illustrate how workflow configurations create a large space of execution plans where each plan induces different accuracy, latency, and cost trade-offs. We then examine two existing approaches to estimating configuration accuracy and their limitations.

\subsection{Compound AI Deployment Optimization}

\begin{figure*}[t]
  \centering
  \includegraphics[width=0.9\textwidth]{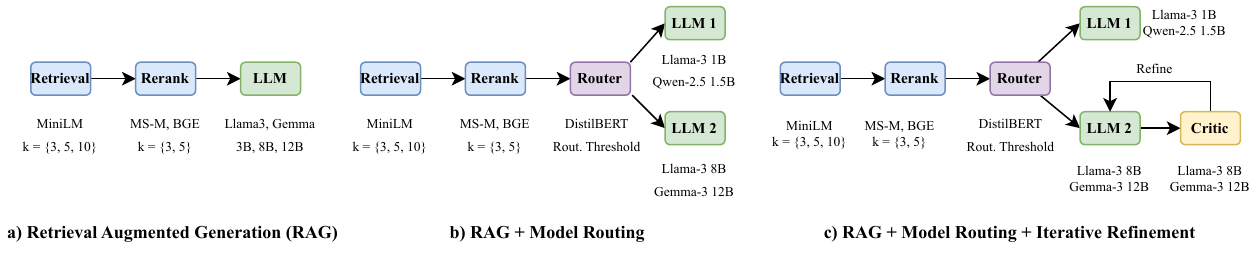}
  \caption{Representative compound AI workflows with illustrative configuration counts.}
  \label{fig:motivation-workflows}
\end{figure*}

To illustrate the deployment optimization problem, we consider three compound AI workflows of increasing complexity, shown in Figure~\ref{fig:motivation-workflows}. The RAG pipeline (a) retrieves documents, reranks them, and feeds the result to a language model that generates an answer~\cite{RAG, Jiang2025RAGO, Ray2024RAGServe}. Each stage exposes model variants and hyperparameters: three retrieval depths, two reranker models at two reranking depths, and six generator models. A workflow configuration assigns one model variant to each stage, producing 72 configurations. Not all queries require a large generator, so a router (b) can dispatch inputs to a small or large generator based on estimated difficulty, expanding the space to 288 configurations~\cite{HybridLLM, ROUTELLM, QLRouter, hu2024routerbenchbenchmarkmultillmrouting, Liu2025CARROT, Sikeridis2024PickLLM}. As smaller models typically produce lower-quality answers, a feedback loop (c) can trigger critic-driven revision, expanding the space to 1728 configurations~\cite{chen2025llmselector, chen2024llmcallsneedscaling, Madaan2023SelfRefine}. Each added stage multiplies the number of configurations that the optimizer must evaluate, and different configurations induce different accuracy, latency, and cost trade-offs. To deploy a selected configuration, an execution plan maps each selected model onto a cluster tier, adding hardware placement to the variant assignment.

For each configuration, the optimizer must estimate how the corresponding execution plan will perform before deployment. System metrics are tractable. End-to-end latency, for example, can be composed from per-stage profiles: a configuration with MiniLM retriever (12 ms), MS-MARCO reranker (38 ms), and Llama-3.1 8B generator (385 ms) yields approximately 435 ms on an RTX 4090. Swapping the generator to Llama-3.2 1B reduces it to roughly 200 ms. Per-stage latencies can be profiled once per (variant, tier) pair and reused across configurations. Cost, throughput, and memory behave similarly.

However, estimating accuracy across configurations requires a different approach. For example, two RAG configurations differing only in the reranker (MS-MARCO vs. BGE) can surface different passages as input for the downstream generator. The same Llama-3.1 8B generator can answer correctly given strong evidence from one reranker and fail given weaker evidence from the other, even though no generator parameter changed. This means that accuracy at one stage depends not only on the model selected there but also on upstream selections. Workflow accuracy is thus a property of the full configuration and cannot be trivially composed from isolated workflow stages. Still, the optimizer must estimate workflow accuracy in order to rank candidate plans and select the best one for deployment.

\subsection{End-to-end accuracy profiling}

The most direct way to estimate configuration accuracy is to profile complete configurations end to end. Each candidate configuration is executed on a representative evaluation dataset and the final workflow output is scored. This captures interactions between stages as they occur in the deployed workflow and provides a faithful reference for plan selection.

\begin{figure}[h]
  \centering
  \includegraphics[width=0.6\columnwidth]{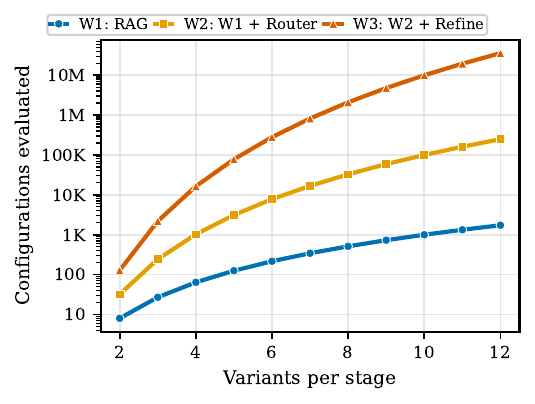}
  \caption{Simulated end-to-end profiling cost for the workflows in Figure~\ref{fig:motivation-workflows}.}
  \label{fig:e2e-profiling-growth}
\end{figure}

Exhaustive end-to-end profiling captures stage interactions directly but requires evaluating every candidate configuration. Figure~\ref{fig:e2e-profiling-growth} shows a simulated profiling cost for the three workflows described in Figure~\ref{fig:motivation-workflows}, measuring the number of complete configuration evaluations required as the number of variants per stage increases. Even for the simplest linear RAG pipeline, the number of configurations that must be evaluated grows into the thousands with only a handful of variants per stage. For the composed workflow with a router and feedback loop, the count exceeds one million. Adding a single variant at any stage creates new combinations with every existing variant in the rest of the workflow, and adding a new stage multiplies the configuration count entirely. The profiling cost of exhaustive end-to-end measurement therefore becomes intractable as the configuration space grows, even for workflows of moderate depth.




\subsection{Product Based Accuracy Estimation}

A natural alternative to exhaustive end-to-end profiling is to estimate configuration accuracy from per-stage measurements. The pipeline accuracy score (PAS), used by IPA~\cite{IPA}, is representative of this approach. For a configuration $c$, PAS assigns one standalone accuracy value to each selected stage variant and combines them multiplicatively,

\begin{equation}
    \text{PAS}(c) = \prod_{s \in S} a(s, v_s),
    \label{eq:pas}
\end{equation}

where $a(s, v_s)$ denotes the standalone accuracy of variant $v_s$ at stage $s$. Each stage can be measured independently and scores can be reused across configurations, making PAS cheap to compute. However, PAS treats stage accuracy contributions as independent scalar factors, ignoring the dependencies between stages, which could lead to poor accuracy estimates.

To test this assumption, we apply PAS to the three workflows in Figure~\ref{fig:motivation-workflows} and compare resulting configuration rankings against measured end-to-end accuracy. Table~\ref{tab:motivation-pas} reports the results. The results show that ranking quality degrades consistently with workflow complexity. Spearman correlation falls from 0.295 on the RAG pipeline to 0.068 on the full composed workflow, with zero Top-5 overlap across all three workflows. 

\begin{table}[t]
  \centering
  \scriptsize
  \caption{PAS-naive ranking against measured end-to-end accuracy.}
  \label{tab:motivation-pas}
  \begin{tabular}{lrrrr}
    \toprule
    Workflow & $\rho$ & $\tau$ & Top-5 & Regret \\
    \midrule
    RAG & 0.295 & 0.203 & 0.00 & +0.322 \\
    RAG + Router & 0.186 & 0.171 & 0.00 & +0.720 \\
    RAG + Router + Loop & 0.068 & 0.023 & 0.00 & +0.657 \\
    \bottomrule
  \end{tabular}
\end{table}

A deployment optimizer relying on PAS can therefore select a plan that satisfies constraints but delivers substantially lower accuracy than alternatives, as reflected by the regret values in Table~\ref{tab:motivation-pas}. The question is whether configuration accuracy can be estimated with enough fidelity to preserve configuration rankings, without requiring full end-to-end measurement.

\section{Atlas Framework Overview}
\label{sec:framework-overview}

\begin{figure*}[t]
  \centering
  \includegraphics[width=\linewidth]{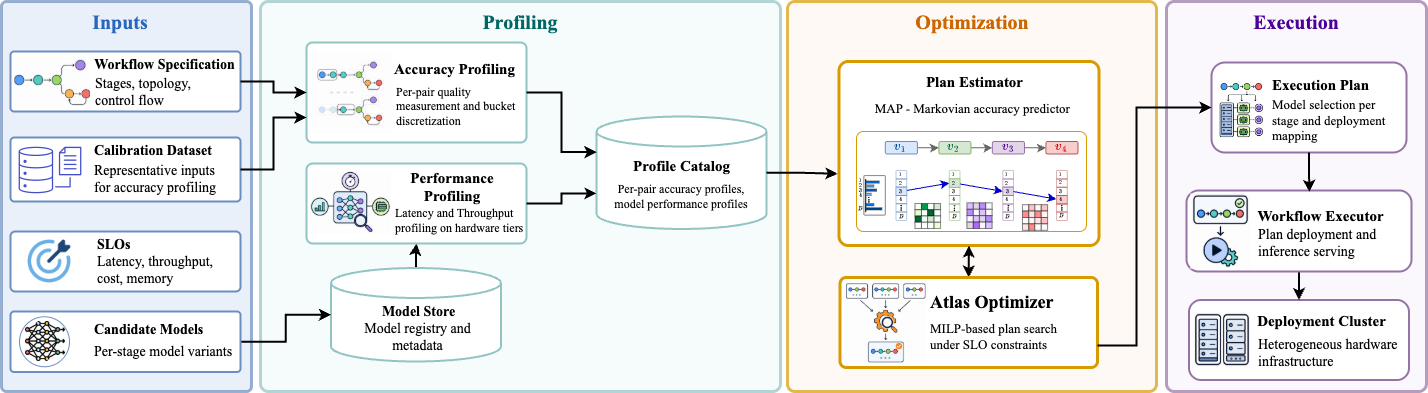}
  \caption{Atlas framework overview: profiling, optimization, and execution phases producing an accuracy-optimized Compound AI execution plan.}
  \label{fig:framework}
\end{figure*}

Atlas is a framework for deployment optimization of compound AI workflows on heterogeneous clusters. It takes a workflow specification, candidate model variants, a calibration dataset, and SLOs as input, and produces an execution plan that assigns one model variant and one hardware placement to each workflow stage. The framework operates in three phases: profiling, optimization, and execution, as shown in Figure~\ref{fig:framework}.

The candidate model variants are registered in the model store, which represents the ecosystem of models available for optimization across workflow stages. The SLOs define the operating constraints under which Atlas selects execution plans, covering latency, throughput, memory, and cost requirements.

In the profiling phase, two independent profilers operate in parallel. The accuracy profiler runs each pair of adjacent stages on the calibration dataset, measuring how the output quality of an upstream stage shifts the quality distribution of its downstream neighbor. This produces per-pair accuracy profiles that keep profiling cost proportional to the number of stage pairs rather than the number of full configurations. The performance profiler draws candidate variants from the model store and measures the latency, throughput, and memory footprint of each variant on each hardware tier. Both sets of profiles are stored in the profile catalog.

In the optimization phase, the plan estimator consumes the per-pair accuracy profiles and applies Markovian Accuracy Predictor (MAP) to estimate end-to-end accuracy for every candidate configuration. The Atlas optimizer selects the model variant assignment and hardware placement per stage that maximizes predicted accuracy subject to the SLO constraints. 

The resulting execution plan is submitted to the workflow executor, which deploys the selected variants on their assigned hardware tiers and serves inference requests until a new plan replaces it.




\subsection{System Model}
\label{sec:system-model}

A compound AI workflow is represented as a directed graph $G = (S, E)$, where each stage $s \in S$ is one AI model or component invocation, and each edge $(s_i, s_j) \in E$ denotes the data dependence between two stages. Each stage exposes a finite set of variants $V_s$. A variant specifies the AI model used at that stage and the model-specific hyperparameters exposed to optimization. A workflow configuration $c$ represents one variant setting for every stage of the workflow. 

Atlas deploys the workflow on a heterogeneous cluster located within one physical site. The cluster contains a finite set of hardware classes $T$, such as CPU workers and GPU workers with different accelerators. 

An execution plan extends a workflow configuration with deployment decisions. Given a configuration $c$, the plan selects, for each stage $s$, the hardware class $t_s$ on which the selected variant $c(s)$ runs:
\begin{equation}
p = \bigl(c,\, (t_s)_{s \in S}\bigr), \quad t_s \in T.
\label{eq:execution-plan}
\end{equation}

The optimizer therefore chooses both the workflow configuration and the
resources used to serve it.

For each stage, variant, and hardware class tuple $(s,v,t)$, the system profiler provides tail execution latency $\ell({s,v,t})$, sustained service capacity $\theta({s,v,t})$, and memory footprint $\mu({s,v,t})$. Each hardware class $t \in T$ has aggregate memory capacity $M_t$ and hourly worker cost $r_t$. These profiles define the system cost of deploying a selected workflow configuration.

Under this model, variant choices determine task accuracy, while placement determines serving behavior. Atlas therefore treats workflow accuracy as a function of the workflow configuration. As the Atlas optimizer relies on the accuracy estimation to rank candidate execution plans, we denote predicted accuracy for an execution plan $p$ extending configuration $c$  as $\hat{E}(p)=\hat{E}(c)$. In Atlas, this is computed through MAP from the local accuracy profiles, further described in Section~\ref{sec:accuracy-model}.

Given an execution plan $p$, Atlas estimates serving feasibility from the system profiles. As workflows can incur different topologies, latency is evaluated over feasible request paths. The workflow specification and the selected control-flow parameters in configuration $c$ define a finite set of paths $\mathcal{P}(c)$. A path $q \in \mathcal{P}(c)$ is a sequence of stage invocations. This means that routed topologies appear as different paths per branch and iteration topologies appear as repeated stage invocations.

Atlas performs offline execution plan selection for a single site cluster. Additionally, we consider that all workers communicate through the same cluster network fabric. Thus, we assume that intra-cluster communication incurs negligible overhead relative to total execution time. Consequently, the system model does not represent inter-stage communication explicitly and attributes end-to-end latency only to stage executions. With this in mind, we model the estimated workflow latency as:

\begin{equation}
L(p)=
\max_{q \in \mathcal{P}(c)}
\left(
\sum_{i=1}^{|q|}
\ell(s_i, c(s_i), t_{s_i})
\right).
\label{eq:latency-model}
\end{equation}

Let $\lambda$ denote the operator-specified workflow throughput target. Atlas uses conservative per-stage provisioning, requiring every deployed stage to sustain the full workflow rate. Since the selected deployment for stage $s$ provides capacity $\theta({s,c(s),t_s})$, throughput feasibility requires:

\begin{equation}
\theta(s, c(s), t_s) \geq \lambda,
\qquad \forall s \in S.
\label{eq:throughput-model}
\end{equation}

Memory feasibility is enforced per hardware class. Let $M_t$ denote the aggregate memory capacity available on workers of class $t$. The total memory footprint of all stages placed on that class must fit within this capacity:

\begin{equation}
\sum_{s \in S: t_s=t}
\mu(s, c(s), t_s)
\leq M_t,
\qquad \forall t \in T.
\label{eq:memory-model}
\end{equation}

Let $r_t$ denote the hourly vendor price of one worker on
hardware class $t$. The hourly deployment cost of execution plan $p$ is

\begin{equation}
C(p)=
\sum_{s \in S} r_{t_s}.
\label{eq:cost-model}
\end{equation}

Equations~\eqref{eq:latency-model}--\eqref{eq:cost-model} define the system-level constraints under which workflow accuracy is optimized.

\subsection{Problem Definition}
\label{sec:problem-definition}

Given a workflow graph $G=(S,E)$, per-stage variant sets
$\{V_s\}_{s\in S}$, hardware classes $T$, system profiles
$(\ell,\theta,\mu)$, hardware capacities and costs $(M,r)$, and operator constraints $(L_{\max},C_{max})$, Atlas selects an execution plan that maximizes predicted task accuracy under latency and cost constraints. The optimization problem is

\begin{equation}
\begin{aligned}
\max_{p} \quad & \hat{E}(p) \\
\text{s.t.} \quad
& L(p) \leq L_{\max}, \\
& C(p) \leq C_{max}
\end{aligned}
\label{eq:optimization-problem}
\end{equation}

The optimization is additionally subject to the throughput and memory feasibility constraints in Equations~\eqref{eq:throughput-model} and~\eqref{eq:memory-model}.

The remaining question is how to construct $\hat{E}(c)$ so that it
preserves the ordering of candidate configurations well enough to support optimization. The next section addresses this question through an accuracy predictor over stage interactions.

\section{MAP: Markovian Accuracy Predictor}
\label{sec:accuracy-model}

Atlas uses the Markovian Accuracy Predictor (MAP) to estimate
end-to-end configuration accuracy from per-pair quality profiles, avoiding exhaustive profiling. MAP tracks intermediate output quality as a discrete state across workflow stages and composes local conditional quality transitions according to the workflow topology. The Plan Estimator invokes MAP to score every candidate configuration before the optimizer selects an execution plan.

\subsection{Workflow and Quality States}
\label{sec:map-quality-states}

MAP assigns every stage output to a quality bucket. Quality refers to the intermediate per-stage signal, such as retriever hit rate or answer F1, that propagates through the workflow and determines workflow accuracy at the terminal stage. MAP discretizes this signal into $B$ buckets whose boundaries are computed per stage from calibration data pooled over variants, making buckets comparable across variants at the same stage.

MAP assumes that the bucketed quality state preserves the information needed to predict downstream quality. Specifically, that it retains the signal relevant for ranking configurations, not all semantic properties of an intermediate output.

A bucket trajectory is the sequence of quality buckets produced as an input moves through the workflow,

\[
    \mathbf{b} = (b_1,\ldots,b_S),
\]
where $b_s \in \{1,\ldots,B\}$ is the bucket realized at stage $s$. For a configuration $c$, MAP induces a trajectory distribution $P(\mathbf{b}\mid c)$. The predicted end-to-end accuracy is

\begin{equation}
    \hat{E}(c)
    =
    \mathbb{E}_{\mathbf{b}\sim P(\mathbf{b}\mid c)}
    [\phi(\mathbf{b},c)],
    \label{eq:map-expected-quality}
\end{equation}

where $\phi$ maps the terminal quality bucket to a scalar accuracy value.

\subsection{Local Transition Model}
\label{sec:map-local-transition}

MAP models quality propagation as a first-order Markov chain over quality buckets, estimating transition probabilities from per-pair quality profiles of adjacent stages. For adjacent stages $s-1$ and $s$, the transition
\begin{equation}
    G_s(b' \mid b,\, v,\, v'), 
    \quad v \in V_s,\; v' \in V_{s-1},
    \label{eq:map-transition}
\end{equation}
gives the probability that stage $s$, using variant $v$, produces quality bucket $b'$ given that stage $s-1$, using variant $v'$, produced quality bucket $b$. Intuitively, $G_s$ captures how the quality produced by an upstream AI model affects the quality distribution of its downstream neighbor. For example, a retriever that retrieves low-quality documents shifts the downstream generator's output distribution towards lower accuracy, regardless of which generator variant is selected.

Each row of $G_s$ fixes one upstream quality bucket $b$ and one variant pair $(v, v')$ and gives a categorical distribution over downstream quality buckets, estimated from calibration samples with smoothing applied to avoid zero-probability transitions in low-support cells. Conditioning only on the immediate upstream bucket and selected variants is a deliberate design choice. It keeps profiling cost local to adjacent stage pairs while still capturing the direct quality interactions that determine configuration rankings.

\begin{figure}[t]
    \centering
    \includegraphics[width=0.85\linewidth]{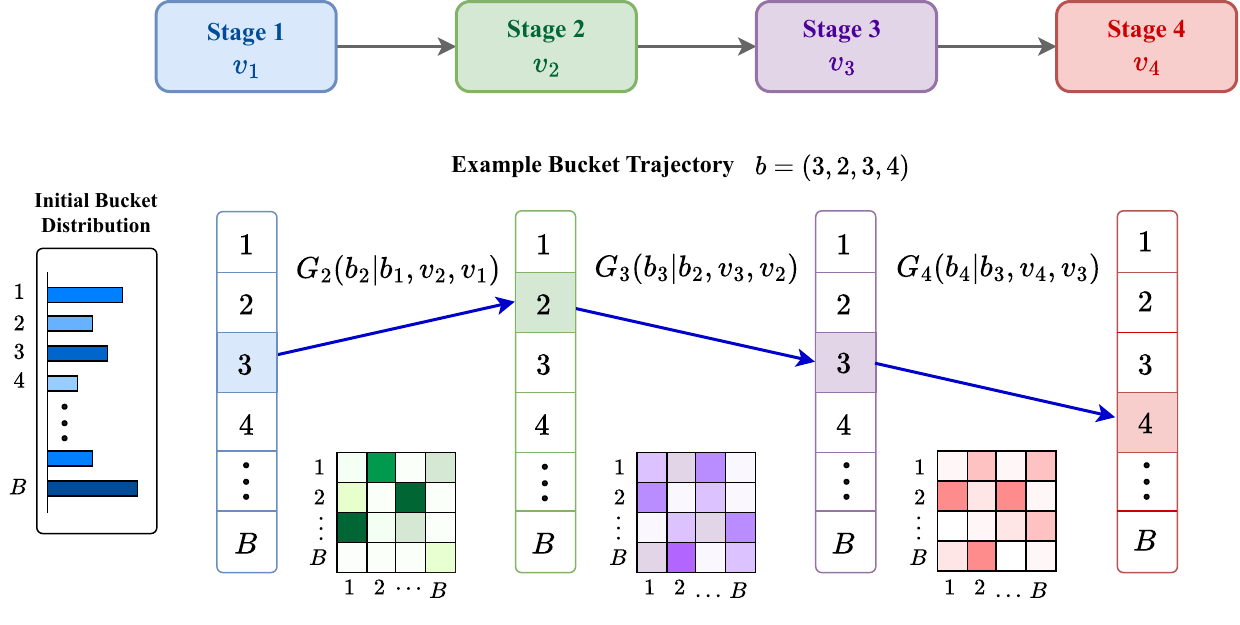}
    \caption{
    MAP's local transition model for a linear workflow. Stage outputs are mapped to quality buckets, adjacent stages define conditional transition matrices, and the highlighted path shows one possible bucket trajectory.
    }
    \label{fig:map-transition}
\end{figure}

As the first workflow stage doesn't have an upstream bucket, MAP estimates the initial distribution 

\begin{equation}
    \pi_{1,b}(v), \quad v \in V_1,
    \label{eq:map-initial-distribution}
\end{equation}

the probability that stage $s=1$ produces quality bucket $b$ when using variant $v$, estimated by running each candidate variant on the calibration inputs and recording the bucket frequency. Figure~\ref{fig:map-transition} shows the  local transition model. An initial quality bucket distribution at stage $s=1$ and a conditional transition matrix $G_s$ for each subsequent stage pair, together covering every candidate variant assignment in the workflow.

\subsection{Topology Operators}
\label{sec:map-topology}

MAP instantiates Equation~\ref{eq:map-expected-quality} for three main topology primitives that cover the most common compound AI workflows: sequential pipelines, routed workflows, and loops. Each primitive defines how quality bucket distributions are propagated through its stages and how the terminal distribution is mapped to a predicted accuracy value. Composed workflows are handled by chaining these primitives, as described in Section~\ref{sec:map-composition}.

\subsubsection{Linear Pipelines}

A linear pipeline executes stages sequentially, where the output of stage $s-1$ becomes the input to stage $s$. Under the local transition model, the trajectory distribution factorizes as

\begin{equation}
    P(\mathbf{b}\mid c)
    =
    \pi_{1,b_1}(v_1)
    \prod_{s=2}^{S}
    G_s(b_s \mid b_{s-1}, v_s, v_{s-1}),
    \label{eq:map-linear-factorization}
\end{equation}

and the predicted accuracy is the expected quality of the terminal bucket,

\begin{equation}
    \hat{E}(c)
    =
    \sum_{\mathbf{b}}
    P(\mathbf{b}\mid c)\, q_{S,b_S},
    \label{eq:map-linear-quality}
\end{equation}

where $q_{S,b_S}$ is the average task accuracy of calibration inputs that produced quality bucket $b_S$ at the terminal stage, obtained from the same calibration data used to estimate the transition matrices.

\subsubsection{Routed Workflows}

A routed workflow sends each input to one of several downstream
branches based on a discretized routing signal $z$, such as an
upstream quality bucket. MAP estimates two quantities for each
candidate router variant $v_R \in V_R$ and branch variant
$v_j \in V_j$. The routing probability is

\begin{equation}
    R_j(z \mid v_R),
    \quad v_R \in V_R,
    \label{eq:map-router-prob}
\end{equation}

For each fixed routing state $z$, $R_j(z \mid v_R)$ gives the probability that router variant $v_R$ sends the input to branch $j$, with $\sum_{j \in \mathcal{J}} R_j(z \mid v_R) = 1$. The branch transition

\begin{equation}
    G_j(b' \mid z, v_R, v_j),
    \quad v_j \in V_j,
    \label{eq:map-branch-transition}
\end{equation}

gives the probability that branch $j$, using variant $v_j$, produces quality bucket $b'$ given routing state $z$ and router variant $v_R$. Intuitively, $R_j$ captures which branch an input is sent to, while $G_j$ captures what quality that branch produces once selected.

Let $\mu^{in}(z)$ denote the distribution over routing states at the router input, estimated from calibration data for a standalone router or supplied by the preceding segment in a composed workflow. The joint trajectory probability is
\begin{equation}
    P(z,j,b' \mid c)
    =
    \mu^{in}(z)\,
    R_j(z \mid v_R)\,
    G_j(b' \mid z, v_R, v_j),
    \label{eq:map-routing-factorization}
\end{equation}
and the predicted accuracy is
\begin{equation}
    \hat{E}(c)
    =
    \sum_{z,j,b'}
    P(z,j,b' \mid c)\, q_{j,b'},
    \label{eq:map-routing-quality}
\end{equation}
where $q_{j,b'}$ is the average accuracy of calibration inputs that were sent to branch $j$ and produced quality bucket $b'$.

\subsubsection{Feedback Loop}

A feedback loop alternates between a producer stage, which generates an output, and an evaluator stage, which scores it and triggers revision if the quality is insufficient (e.g., a generator and critic in a self-refinement workflow). The producer's initial output follows the bucket distribution $\pi_{0,b}(v_p)$ defined in Section~\ref{sec:map-local-transition}. The feedback transition

\begin{equation}
    T(b' \mid b, v_p, v_e),
    \quad v_p \in V_p,\; v_e \in V_e,
    \label{eq:map-refinement-transition}
\end{equation}

gives the probability that one feedback step moves the output from quality bucket $b$ to quality bucket $b'$ under producer $v_p$ and evaluator $v_e$. Intuitively, $T$ captures how effectively an evaluator variant drives quality improvements in the producer output across feedback steps.

For a fixed budget $K$, the quality distribution after $K$ feedback steps is

\begin{equation}
    \mu_K = \pi_0(v_p)\, T(v_p, v_e)^K,
    \label{eq:map-refinement-state}
\end{equation}

and the predicted accuracy is

\begin{equation}
    \hat{E}(c)
    =
    \sum_{b=1}^{B}
    \mu_{K,b}\, q_b,
    \label{eq:map-refinement-quality}
\end{equation}

where $q_b$ is the average task accuracy of calibration inputs that produced quality bucket $b$ after $K$ feedback steps.

The fixed budget model assumes every input undergoes exactly $K$
feedback steps. When the workflow stops adaptively based on evaluator score, MAP augments $T$ with an absorbing stop state,

\begin{equation}
    \tilde{T}
    =
    \begin{bmatrix}
        T_{\mathrm{cont}} & p_{\mathrm{stop}} \\
        0 & 1
    \end{bmatrix},
    \label{eq:map-absorbing-chain}
\end{equation}
where $T_{\mathrm{cont}}$ captures transitions that continue the
feedback loop and $p_{\mathrm{stop}}$ gives the stopping probability from each quality bucket. The augmented chain produces two outputs: the final quality distribution, used to compute $\hat{E}(c)$, and the expected iteration count per input, which Atlas uses to estimate latency and throughput for the execution plan.

\subsection{Workflow Operator Composition}
\label{sec:map-composition}

A composed workflow chains multiple topology primitives into a single workflow. For example, a sequential RAG pipeline may feed into a feedback loop backend. MAP treats each segment $m$ as an operator

\begin{equation}
    F_m(\mu^{in}_m, c_m),
    \label{eq:map-segment-operator}
\end{equation}

which maps an input quality bucket distribution $\mu^{in}_m$ and
segment configuration $c_m$ to an output quality bucket distribution $\mu^{out}_m$. Each segment applies its own topology operator from Section~\ref{sec:map-topology}.

Passing quality information across segment boundaries requires one additional profiled quantity. Each segment defines its own bucket boundaries from calibration data pooled over its own stages, so the output bucket distribution $\mu^{out}_m$ of segment $m$ is not directly comparable to the input bucket space of segment $m+1$. Thus, the boundary transition

\begin{equation}
    G_{\mathrm{entry}}(b' \mid b, v_{\mathrm{entry}}),
    \quad v_{\mathrm{entry}} \in V_{\mathrm{entry}},
    \label{eq:map-entry-transition}
\end{equation}

gives the probability that the entry stage of segment $m+1$ produces quality bucket $b'$ given that segment $m$ ended in quality bucket $b$. The entry distribution of segment $m+1$ is 

\begin{equation}
    \mu_{m+1,0}(b')
    =
    \sum_{b=1}^{B}
    \mu^{out}_m(b)\,
    G_{\mathrm{entry}}(b' \mid b, v_{\mathrm{entry}}),
    \label{eq:map-boundary-update}
\end{equation}

after which segment $m+1$ applies its own topology operator,

\begin{equation}
    \mu^{out}_{m+1}
    =
    F_{m+1}(\mu_{m+1,0}, c_{m+1}).
    \label{eq:map-composed-output}
\end{equation}

This preserves cross-segment quality dependence without conditioning on the full upstream trajectory, keeping composition profiling cost additive across segment boundaries.

\subsection{Complexity Analysis}
\label{subsec:profiling-cost}

\subsubsection{Profiling cost} MAP profiles only adjacent stage pairs. A linear segment of $S$ stages requires $O(SV^2B)$ conditional rows, where $V$ is the maximum variants per stage and $B$ the number of quality buckets. A routed segment adds $O(V_R B_z \sum_{j} |V_j|)$ rows for the router and branches, while a feedback loop adds $O(V_p V_e B)$ for the feedback transition. Each segment boundary adds $O(V_{\mathrm{entry}}B)$. Overall, profiling cost grows additively across primitives and boundaries, compared with $O(V^S)$ for exhaustive end-to-end profiling.

MAP's local structure reduces the cost of workflow evolution. Adding one variant to a stage requires $O(V \cdot B)$ new pairwise transitions instead of $V^{S-1}$ new configurations for end-to-end profiling. Inserting a new stage requires profiling only its two adjacent stages instead of complete re-profiling.

\subsubsection{Prediction cost} For each topology, MAP operates on bucket distributions directly. A linear segment requires one matrix-vector multiplication over $B$ buckets per stage, giving a prediction cost of $O(SB^2)$ per configuration. For routed segments, MAP sums over branch assignments. For feedback loops, it sums over iteration counts. In both cases prediction cost remains polynomial in $B$ and the number of stages.
\section{Plan Optimizer}
\label{sec:optimizer}

The problem in Equation (7) selects a workflow configuration and deployment tier per stage to maximize accuracy under SLO constraints on a heterogeneous cluster, as defined in Section~\ref{sec:system-model}. We cast it as a MILP using two structural properties. Predicted accuracy $\{\hat{E}(c)\}$ from MAP is precomputed per configuration and enters the objective as a constant coefficient, while the system constraints are linear in the deployment indicators.

\textbf{Decision variables.}
Let $\mathcal{C} = \prod_{s \in S} V_s$ denote the configuration space. The MILP uses the configuration selector $\xi_c \in \{0,1\}$ with $\sum_{c} \xi_c = 1$; and deployment indicators $\omega_{s,v,t} \in \{0,1\}$ with $\sum_{v,t} \omega_{s,v,t} = 1$ for each $s$, denoting that stage $s$ runs variant $v$ on tier $t$. Variant selection is the projection $x_{s,v} := \sum_t \omega_{s,v,t} = \sum_{c\,:\,c(s)=v} \xi_c$, linking configuration selection to deployment.

\textbf{Latency.}
Each request path through the workflow must satisfy the latency SLO:
\begin{equation}
\sum_{s \in q} \sum_{v,t} \ell(s,v,t)\, \omega_{s,v,t} \leq L_{\max}, \quad \forall q \in \mathcal{P}.
\label{eq:milp-latency}
\end{equation}

\textbf{Throughput.}
Each deployed stage must serve its effective request rate:
\begin{equation}
\sum_{v,t} \theta(s,v,t)\, \omega_{s,v,t} \ge K_s \lambda
\quad \forall s \in S,
\label{eq:milp-throughput}
\end{equation}
where $K_s$ is the iteration bound of the feedback loop enclosing $s$, and $K_s = 1$ outside loops. Thus, a stage inside a loop serves each request up to $K_s$ times. This mirrors the latency constraint, where loop iterations are unrolled into the request paths $\mathcal{P}$.

\textbf{Memory.}
The aggregate footprint hosted on each tier must fit its capacity:
\begin{equation}
\sum_{s,v} \mu(s,v,t)\, \omega_{s,v,t} \le M_t
\quad \forall t \in \mathcal{T}.
\label{eq:milp-memory}
\end{equation}

\textbf{Cost.}
Hourly deployment cost sums each stage's tier rate:
\begin{equation}
\sum_{s,v,t} r_t\, \omega_{s,v,t} \le C_{max}.
\label{eq:milp-cost}
\end{equation}

\textbf{Objective.}
Predicted end-to-end accuracy enters as a precomputed coefficient on
$\xi_c$:
\begin{equation}
\max_{\xi,\, \omega}
\;\;
\sum_{c \in \mathcal{C}} \hat{E}(c)\, \xi_c .
\label{eq:milp-objective}
\end{equation}

The values $\{\hat{E}(c)\}_{c \in \mathcal{C}}$ are predicted by MAP from the local pairwise profiles and stored before the solver is invoked, keeping the program linear in all decision variables. A small cost-minimizing tiebreaker $\varepsilon \sum_{s,v,t} r_t \, \omega_{s,v,t}$ with $\varepsilon = 10^{-4} / C_{\max}$ is subtracted from the objective to prefer cheaper placements among configurations with equal predicted accuracy. 

\textbf{Complexity.}
All constraints and the objective are linear, so the LP relaxation is convex. Integrality of $\xi$ and $\omega$ makes MILP solving NP-hard in general. MAP reduces profiling cost, while the MILP still represents the configuration space through $\xi_c$. Because MAP scores configurations before solving, those with lower predicted accuracy at equal resource footprint can be pruned to keep the problem tractable.


\begin{table*}[t]
    \centering
    \scriptsize
    \setlength{\tabcolsep}{3pt}
\caption{Workflows used in the evaluation.}
\label{tab:evaluation-workflows}
\begin{tabular}{p{0.055\textwidth} p{0.19\textwidth} p{0.17\textwidth} p{0.48\textwidth} r}
\toprule
ID & Topology & Workflow & Configuration space & Configs. \\
\midrule
W1 & Linear &
RAG &
$k_r \in {5,10,20}$; reranker $\in {\text{MS-MARCO}, \text{BGE-base}}$; $k_{rr} \in {3,5}$; generator $\in$ Llama-3.2 1B/3B, Llama-3.1 8B, Gemma-3 1B/4B/12B, Phi-3 3.8B, Qwen2.5 1.5B~\cite{minilm, MSMARCO, gemma2025gemma3, grattafiori2024llama3, faiss} &
96 \\
\midrule
W2 & Linear + Routing &
RAG + Router &
W1 with fixed $k_r$ and $k_{rr}$; reranker $\in {\text{MS-MARCO}, \text{BGE-base}}$; DistilBERT threshold $t \in {0.3,0.4,0.5,0.6}$; local generator $\in$ Llama-3.2 1B/3B, Gemma-3 1B, Qwen2.5 1.5B; remote generator $\in$ Llama-3.1 8B, Gemma-3 4B/12B, Phi-3 3.8B. &
128 \\
\midrule
W3 & Linear + Feedback Loop &
RAG + Refine &
Five selected W1 configurations; four producer and evaluator pairs drawn
from Llama-3.2 1B/3B and Gemma-3 4B producers and Llama-3.2 1B and
Gemma-3 4B evaluators; feedback budget $K \in \{1,3,5\}$. &
60 \\
\midrule
W4 & Linear + Routing + Feedback Loop &
RAG + Router + Refine &
W1 with fixed $k_r$ and $k_{rr}$; reranker $\in {\text{MS-MARCO}, \text{BGE-base}}$; DistilBERT threshold $t \in {0.3,0.4,0.6}$; local producer $\in$ Llama-3.2 1B/3B, Gemma-3 4B; evaluator $\in$ Llama-3.2 1B, Gemma-3 4B; remote generator $\in$ Gemma-3 4B/12B, Llama-3.1 8B, Phi-3 3.8B, Mistral 7B, Qwen2.5 7B; $K=5$. &
216 \\
\bottomrule
\end{tabular}
\end{table*}

\section{Evaluation}
\label{sec:evaluation}

This section presents a series of experiments as means to evaluate Atlas. Section~\ref{sec:eval:setup} details the carried-out experiments, experimental frameworks, and evaluation objectives, while  Sections~\ref{sec:prediction-quality} through~\ref{sec:bucket} present the results.

\subsection{Experimental Setup}
\label{sec:eval:setup}

Atlas is implemented in Python and published as an open-source framework within the Polaris project\footnote{\url{https://github.com/polaris-slo-cloud/Atlas}}. For MILP optimization, it uses the CBC solver through the \texttt{python-mip} library~\cite{Forrest2024CBC}. We evaluate Atlas by measuring how well MAP predicts accuracy across different compound AI workflows and whether these predictions translate into effective optimizer decisions. We first assess whether local quality profiles preserve the decision-relevant ordering of configurations across topologies presented in Section~\ref{sec:accuracy-model}. We then test the execution plans produced by the Atlas optimizer against baselines under both homogeneous and heterogeneous cluster settings.

\subsubsection{Workflows and Datasets}

We evaluate Atlas across four workflows that cover the topology patterns modeled by the framework. The linear RAG workflow~\cite{RAG} (W1) tests conditional quality propagation across sequential stages. W2 extends W1 with a learned router~\cite{HybridLLM, ROUTELLM} that dispatches inputs to different generator branches. W3 extends W1 with a generator-critic feedback loop~\cite{Madaan2023SelfRefine, chen2025llmselector}, testing iterative quality evolution. W4 combines routing and refinement on top of W1, testing whether MAP can pass quality distributions across topology boundaries. Table~\ref{tab:evaluation-workflows} summarizes the configuration space for each workflow, including model variants and hyperparameters. W2, W3, and W4 each build on a subset of W1 configurations, extending them with routing, refinement, or both. Each configuration count is the product of the listed factor cardinalities, for example $3 \times 2 \times 2 \times 8 = 96$ for W1.

All workflows are evaluated on the SQuAD~\cite{squadv2} dataset using answer F1 as the end-to-end accuracy metric. Quality labels at intermediate stages describe whether retrieval and reranking preserve the evidence needed by the generator, the router's dispatch decision and downstream branch quality, and the quality trajectory across refinement steps. MAP uses $B = 4$ quality buckets. Bucket boundaries are computed per stage from 200 calibration samples pooled over all variants at that stage. 100 held-out samples are used for evaluation.

\subsubsection{Baselines}
We compare MAP against three accuracy estimation baselines. PAS-naive multiplies standalone benchmark accuracies per stage (Eq.~\ref{eq:pas}), following the Pipeline Accuracy Score used by IPA~\cite{IPA}. PAS-fair is the strongest PAS-style baseline we can construct per topology. It retains scalar composition but replaces standalone scores with topology-aware terms drawn from the same calibration data available to MAP, such as conditioned stage accuracies for linear workflows and routing-weighted branch accuracies for routed workflows. The oracle reference profiles every configuration end to end on the evaluation dataset, following the exhaustive profiling approach used by~\cite{Ahmad2024Loki, Gravara2026Compass}. For optimizer evaluation, we compare execution plans produced by Atlas against IPA and Loki~\cite{Ahmad2024Loki} on a homogeneous cluster, restricting Loki to single-variant-per-stage placement for direct plan-level comparability.

\subsubsection{Infrastructure}

The evaluation uses three hardware tiers: an NVIDIA RTX 4090 (24 GB VRAM, \$0.40/h), an NVIDIA RTX A4000 (16 GB VRAM, \$0.20/h), and 32-core x86 CPU tier (32 GB RAM, \$0.10/h). Hourly tier costs are set to approximate public cloud rates for comparable hardware. Experiments in Sections~\ref{sec:prediction-quality} through \ref{sec:plan-quality-homogeneous} run on a single RTX 4090 server. The heterogeneous evaluation \ref{subsec:heterogeneous} deploys all three tiers on a K3s cluster spanning three physical nodes.

\subsubsection{Metrics}

We use two groups of evaluation metrics. For the accuracy model, we report Spearman and Kendall rank correlation against the oracle ranking, top-5 overlap, and top-1 regret, defined as the accuracy difference between the oracle's best configuration and the one ranked first by the predictor. Rank correlation and top-k overlap evaluate whether the predictor preserves the ordering needed for optimization, while regret measures the practical cost of prediction errors. For the optimizer, we report the measured end-to-end accuracy of the selected execution plan, its p99 latency, and hourly deployment cost under varying SLOs and offered loads.

\begin{table*}[t]
\centering
\small
\caption{MAP prediction quality across workflows compared to PAS and PAS-fair.}
\label{tab:prediction-quality}
\begin{tabular}{llrrrrrr}
\toprule
Workflow & Predictor & $n$ & Spearman $\rho$ & Kendall $\tau$ & MAE & Top-5 overlap & Top-1 regret \\
\midrule
W1 & PAS-naive & 96 & 0.295 & 0.203 & 0.279 & 0.00 & +0.322 \\
W1 & PAS-fair  & 96 & 0.555 & 0.378 & 0.085 & 0.00 & +0.240 \\
W1 & MAP    & 96 & \textbf{0.947} & \textbf{0.805} & \textbf{0.036} & \textbf{0.80} & \textbf{+0.000} \\
\midrule
W2 & PAS-naive & 128 & 0.186 & 0.171 & 0.451 & 0.00 & +0.720 \\
W2 & PAS-fair  & 128 & 0.425 & 0.329 & 0.161 & 0.20 & +0.699 \\
W2 & MAP    & 128 & \textbf{0.921} & \textbf{0.757} & \textbf{0.098} & \textbf{0.80} & \textbf{+0.046} \\
\midrule
W3 & PAS-naive & 60 & $-$0.372 & $-$0.250 & 0.267 & 0.20 & \textbf{+0.000} \\
W3 & PAS-fair  & 60 & 0.619 & 0.487 & 0.126 & 0.40 & \textbf{+0.000} \\
W3 & MAP    & 60 & \textbf{0.783} & \textbf{0.595} & \textbf{0.069} & \textbf{0.60} & +0.127 \\
\midrule
W4 & PAS-naive & 216 & 0.068 & 0.023 & 0.642 & 0.00 & +0.657 \\
W4 & PAS-fair  & 216 & 0.294 & 0.234 & 0.152 & 0.20 & +0.632 \\
W4 & MAP    & 216 & \textbf{0.882} & \textbf{0.697} & \textbf{0.076} & \textbf{0.60} & \textbf{+0.094} \\
\bottomrule
\end{tabular}
\end{table*}

The evaluation targets four objectives. First, it assesses whether MAP preserves the measured ranking of workflow configurations across all topology classes (\ref{sec:prediction-quality}). Second, it measures the profiling cost at which MAP achieves a useful ranking signal relative to exhaustive end-to-end measurement (\ref{sec:profiling-cost}). Third, it evaluates whether the Atlas optimizer, guided by MAP, produces execution plans that match oracle-quality plan selection compared to existing approaches (\ref{sec:plan-quality-homogeneous}). Fourth, it tests whether Atlas's joint optimization over variant selection and heterogeneous tier placement produces better execution plans than single-tier deployment strategies (\ref{subsec:heterogeneous}).

\subsection{MAP Prediction Quality}
\label{sec:prediction-quality}

\begin{figure}[h]
\centering
\includegraphics[width=0.9\columnwidth]{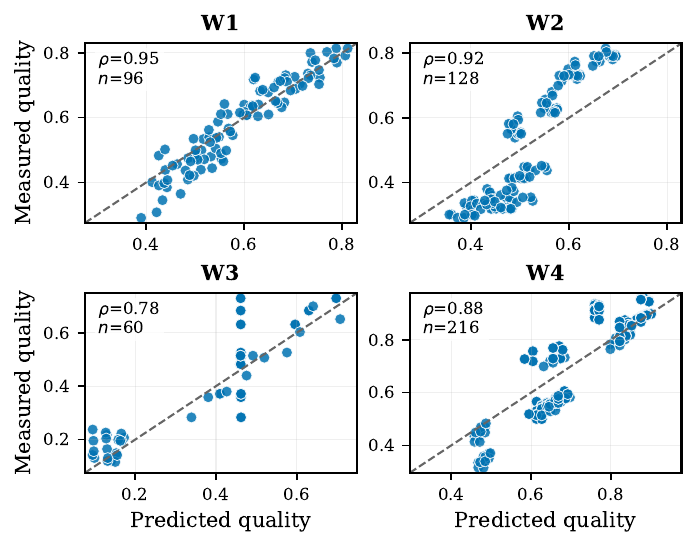}
\caption{MAP predicted versus measured accuracy across four workflows.}
\label{fig:ranking-quality}
\end{figure}

We first evaluate whether MAP preserves the configuration ranking induced by end-to-end measurements. This is the accuracy model's primary requirement for optimization, because the optimizer uses predicted accuracy to compare candidate configurations under serving constraints. Table~\ref{tab:prediction-quality} reports ranking, calibration, and selection metrics for four workflows, and Figure~\ref{fig:ranking-quality} shows MAP's predicted accuracy against measured accuracy for each configuration.


MAP gives the strongest ranking signal on all four workflows, with Spearman correlation ranging from $\rho=0.78$ on W3 to $\rho=0.947$ on W1, while also achieving the best Kendall correlation and lowest regret in the table. Figure~\ref{fig:ranking-quality} confirms this visually. W1 and W2 concentrate close to the diagonal, indicating that MAP is both well ranked and calibrated on linear and routed workflows. W3 and W4 show wider spread because refinement introduces additional variance, but the point clouds still preserve the ordering needed for plan selection.

The wider spread on W3 follows from the first-order Markov assumption. MAP applies one profiled feedback transition $T(v_p, v_e)$ $K$ times, and repeated application mixes toward its stationary distribution, gradually erasing the bucket distribution passed from the upstream W1 segment. Predicted accuracy across configurations that differ only in their W1 segment collapses from a spread of $0.29$ at $K=1$ to at most
$0.03$ at $K \ge 3$, while the measured spread remains up to $0.36$, and MAE grows from $0.036$ to $0.095$. Nevertheless, the ordering within each W1 segment survives, which keeps MAP the best ranking predictor on W3. Conditioning the feedback transition on the quality bucket at loop entry would restore the upstream dependence at a profiling cost multiplied by $B$.

The gap between MAP and PAS-style baselines grows with compositional complexity. On W1, MAP improves Spearman correlation over PAS-fair by 0.39; on W4, the gap increases to 0.59. This pattern reflects that scalar composition becomes less reliable as workflow behavior depends on cross-stage accuracy propagation. PAS-fair can correct some standalone effects, but it does not represent how retrieval accuracy changes router decisions or how routed outputs affect refinement. PAS-naive is anti-correlated on W3 ($\rho=-0.37$), meaning benchmark-product composition can invert the true configuration ordering.

Top-5 overlap and regret metric reinforce this result. MAP recovers 60-80\% of the true top-5 configurations across all workflows, while PAS-naive achieves zero overlap on W1, W2, and W4. Since the optimizer selects from the top-ranked feasible configurations, accurate ranking at the top of the list directly determines the quality of the chosen execution plan.

\subsection{Profiling Cost}
\label{sec:profiling-cost}

\begin{figure}[h]
\centering
\includegraphics[width=\columnwidth]{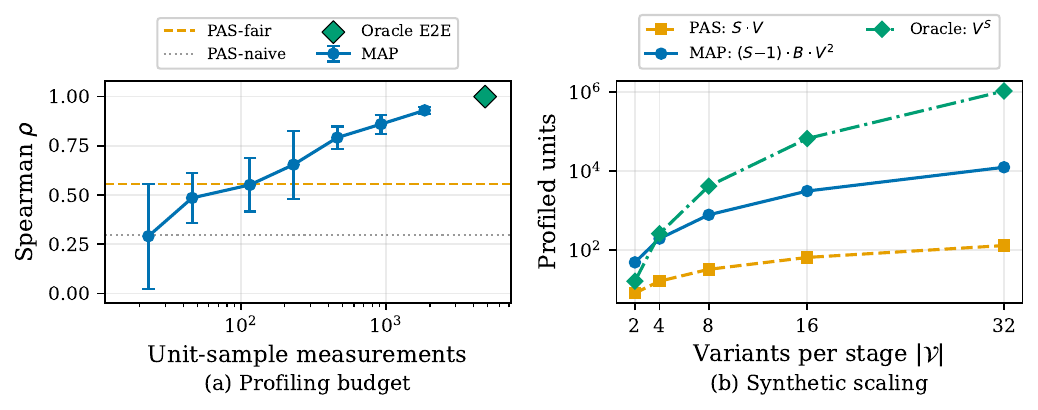}
\caption{Profiling cost of MAP compared to PAS and exhaustive profiling.}
\label{fig:profiling-cost}
\end{figure}

We next evaluate whether MAP reduces the measurement cost needed to obtain a useful ranking signal. Figure~\ref{fig:profiling-cost} compares MAP with PAS baselines and exhaustive end-to-end profiling.

Figure~\ref{fig:profiling-cost}(a) shows Spearman $rho$ as the number of per-pair measurements increases. W1 contains 23 such terms, so using $K$ calibration samples per term requires $23K$ measurements. PAS-fair and PAS-naive appear as horizontal lines because their scalar estimates do not improve with additional workflow-specific measurements. End-to-end profiling gives the reference ranking but requires measuring all 96 configurations on 200 samples. MAP surpasses PAS-fair with 115 measurements and reaches $\rho=0.93$ with 1,840 measurements, recovering most of the reference ranking signal at 2.6$\times$ lower cost.

Figure~\ref{fig:profiling-cost}(b) shows how this gap grows with the number of variants per stage in a synthetic scaling scenario. End-to-end profiling scales with the full configuration space, whereas MAP scales with local pairwise terms. At small variant counts $(|\mathcal{V}| \leq 4)$, end-to-end profiling remains practical and may require fewer measurements than MAP's pairwise terms. However, the gap inverts quickly: at $|\mathcal{V}| = 8$, MAP requires $5.3\times$ fewer profiling units, and at $|\mathcal{V}| = 32$, the gap widens to over $80\times$.

\subsection{Plan Selection on a Homogeneous Cluster}
\label{sec:plan-quality-homogeneous}

We evaluate the execution plans produced by Atlas against those of IPA and Loki on a homogeneous single-tier cluster (RTX 4090, 24 GB memory). We restrict placement to a single hardware tier to match the conditions under which IPA and Loki are designed, and evaluate heterogeneous placement separately in~\ref{subsec:heterogeneous}. Each system is given the same workflow, same candidate variants, and same latency SLO. We report the measured accuracy of the chosen plan.

\begin{figure}[h]
\centering
\includegraphics[width=0.85\columnwidth]{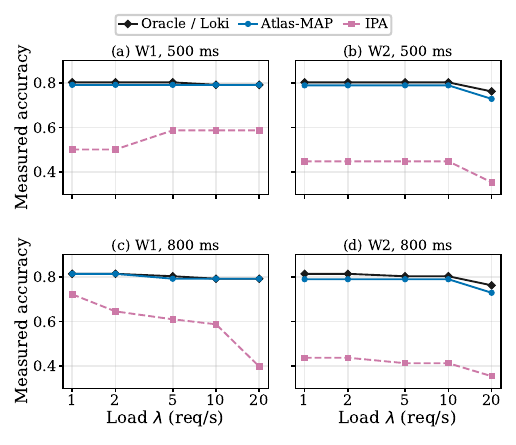}
\caption{Execution plan accuracy comparison with baselines over varying SLOs.}
\label{fig:homogeneous}
\end{figure}

\begin{figure*}[t]
  \centering
  \includegraphics[width=0.9\linewidth]{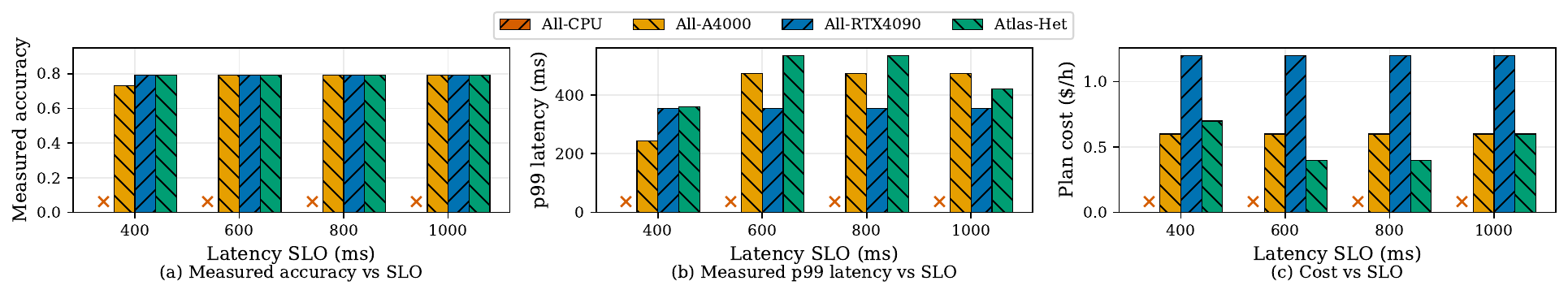}
  \caption{Homogeneous versus heterogeneous placement on W1: measured accuracy, latency, and plan cost across latency SLOs.}
  \label{fig:eval-heterogeneous}
\end{figure*}


To isolate the contribution of the accuracy model, all three systems are run through Atlas's MILP with identical constraints. Under single-variant-per-stage assignment, Loki reduces to a direct end-to-end accuracy lookup and is plotted together with the oracle reference. Loki's full formulation supports concurrent multi-variant deployment per stage, but we restrict to single-variant assignment for direct plan-level comparability. We evaluate on W1 and W2 at two latency SLOs (500 ms, 800 ms) and five offered loads ($\lambda \in \{1, 2, 5, 10, 20\}$ req/s). W3 and W4 are excluded as neither IPA nor Loki express a system model that supports feedback loop topologies.

Figure~\ref{fig:homogeneous} reports measured accuracy of the selected execution plan as a function of offered load under varying SLOs. Atlas stays within 0.03 of Oracle/Loki across all loads and SLO tiers on both workflows, while IPA falls sharply with load, reaching 0.40 on W1 and 0.35 on W2 at $\lambda$ = 20. The accuracy gap between Atlas and IPA grows with workflow complexity, exceeding 0.30 on every W2 load with a median gap of 0.36. This confirms findings from Section~\ref{sec:prediction-quality}, where PAS achieves weak Spearman correlation on routed workflows. Weak accuracy estimation leads to misordered configurations, which in turn leads the optimizer to select worse execution plans. Since all three systems share the same optimizer, the observed differences are attributable solely to the accuracy model.

\subsection{Plan Selection on a Heterogeneous Cluster}
\label{subsec:heterogeneous}

\begin{figure}[h]
  \centering
  \includegraphics[width=0.85\columnwidth]{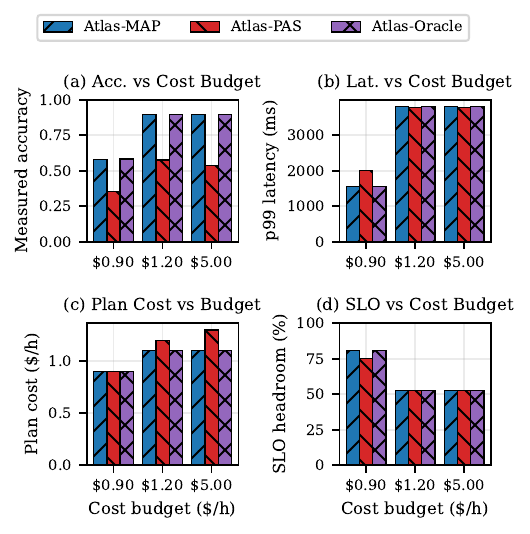}
  \caption{Accuracy predictor comparison on W4 with heterogeneous placement}
  \label{fig:eval-predictor}
\end{figure}

We evaluate whether heterogeneous tier placement produces better execution plans than single-tier strategies. We hold the accuracy predictor fixed to MAP so that observed differences are attributable solely to placement choice. Under four latency SLOs (400--1000~ms), we compare four strategies on W1: All-CPU, All-A4000, All-RTX4090, and Atlas-Het, which uses the full MILP with unrestricted tier assignment per stage.

Figure~\ref{fig:eval-heterogeneous} reports measured accuracy, p99 latency, and plan cost per SLO. All-CPU is infeasible because LLM p99 on CPU exceeds 1.6~s. All-A4000 reaches measured accuracy of 0.73 at the 400~ms SLO, while All-RTX4090 and Atlas-Het reach 0.79, because A4000 latency forces the optimizer to select a smaller LLM variant. At 600~ms and above, all GPU-based strategies converge on 0.79. Atlas-Het achieves the same accuracy as All-RTX4090 at 42\% lower plan cost at 400~ms and 33\% lower at 600 and 800~ms, by placing the embedder and reranker on cpu-edge and the LLM on A4000. This cost reduction follows from the tiebreaker term in the MILP objective, which selects the cheapest placement among equally-accurate configurations. Heterogeneous placement therefore matches single-tier accuracy at lower cost across feasible SLOs.

We next test whether MAP retains oracle-quality plan selection on the heterogeneous cluster for W4. We fix placement to Atlas-Het and vary only the accuracy predictor across Atlas-MAP, Atlas-PAS, and Atlas-Oracle. We sweep three cost budgets (\$0.90, \$1.20, \$5.00) under a loose SLO of 8000~ms to ensure that the cost budget, not latency, is the binding constraint in this experiment.

Figure~\ref{fig:eval-predictor} reports measured accuracy, p99 latency, and plan cost across cost budgets. Atlas-MAP matches Atlas-Oracle at every budget, reaching 0.58 at \$0.90 and 0.90 at \$1.20 and \$5.00. Atlas-PAS lags by 0.22 to 0.35 in measured accuracy, its p99 latencies are approximately 450~ms higher, and exceed Atlas-MAP cost by \$0.20/h at the loose budget. Since placement and optimizer are held fixed, these differences are attributable to the accuracy model. PAS selects bge-base as the reranker because its standalone benchmark score is higher than ms-marco, but bge-base interacts worse with downstream stages. Additionally, at 494~ms p99 on cpu-edge, bge-base cannot be placed on the cheap tier, forcing the MILP onto A4000 and increasing cost. MAP conditions the reranker's contribution on the quality it passes downstream, correctly selects ms-marco, and recovers oracle-quality plans across the full budget range.

\subsection{Bucket Sensitivity Analysis}
\label{sec:bucket}

\begin{figure}[h]
  \centering
  \includegraphics[width=0.8\columnwidth]{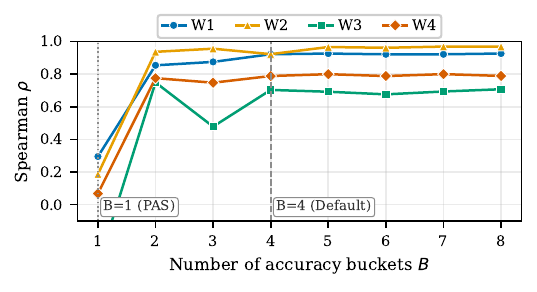}
  \caption{MAP ranking quality as a function of bucket count $B$ across four workflows}
  \label{fig:bucket-sensitivity}
\end{figure}

We analyze MAP's sensitivity to the number of quality buckets $B$, which sets the granularity of intermediate output discretization. Figure~\ref{fig:bucket-sensitivity} shows MAP's ranking quality as a function of the number of $B$ for all four workflows. At $B = 1$, each stage is described by a single scalar accuracy value and MAP reduces to PAS-style composition. Increasing $B$ to 2 produces the largest improvement across all workflows, as even a binary partition allows MAP to condition downstream behavior on upstream quality. All workflows plateau by $B = 4$, with marginal gains beyond that point. The dip at $B = 3$ on W3 occurs because W3's F1 distribution is heavily concentrated at 0 and 1. This causes the equal-frequency quantile boundaries to land on thresholds that merge failures with mediocre outputs into a single bucket, losing the discriminative split that $B = 2$ and $B = 4$ preserve. We use $B = 4$ as the default throughout the evaluation, as it provides stable ranking quality across all topologies. Automatic bucket selection from calibration data is a natural extension of this sweep.

\section{Related Work}
\label{sec:related}

Prior work relevant to Atlas spans three areas: 1) inference serving, 2) accuracy estimation for multi-stage workflows, and 3) optimization formulations for inference deployment.

\subsection{Inference Serving Systems}

Inference serving systems such as Clipper, Clockwork, and Nexus establish the substrate for low-latency ML inference through batching, model management, GPU scheduling, and predictable execution under latency SLOs~\cite{Crankshaw2017Clipper,Gujarati2020Clockwork,Shen2019Nexus}. A second line of work makes accuracy a first-class objective by selecting among model variants under latency, throughput, or cost constraints. ModelSwitching switches to cheaper models under load spikes, INFaaS automates model and hardware selection, Cocktail optimizes ensembles, RAMSIS selects models using inter-arrival-aware scheduling, and Proteus performs accuracy scaling for high-throughput serving~\cite{Zhang2020ModelSwitching,Romero2021INFaaS,Gunasekaran2022Cocktail,
Mendoza2024RAMSIS, Proteus, SLOpt, Hu2023MOSEL, Biathlon}.

These systems laid the foundation for accuracy-aware inference serving but target a single task or model pool at one endpoint. Atlas targets compound AI workflows where downstream accuracy depends on upstream output quality.

\subsection{Accuracy Estimation for AI Workflows}

Two lines of approaches estimate the accuracy of AI workflows. Product-based surrogates compose standalone stage accuracies multiplicatively, of which IPA's Pipeline Accuracy Score (PAS) is
representative~\cite{IPA}. This approach is computationally efficient since per-stage measurements can be reused across configurations, but the independence assumption fails when upstream output quality affects downstream stage behavior. Several approaches rely on end-to-end accuracy profiling on a representative dataset. This faithfully captures inter-stage interactions but scales with the full configuration space~\cite{Ahmad2024Loki,Wu2022JellyBean,
Zhang2024Vulcan,Romero2021Llama, Gravara2026Compass}. As the number of stages, AI model variants, and hyperparameters, grows, exhaustive end-to-end profiling becomes intractable for deployment optimization. 

In Atlas, MAP sits between these two extremes. It profiles local conditional quality transitions between adjacent workflow stages, preserving the cross-stage accuracy dependencies that product-based surrogates discard, while keeping profiling cost proportional to the number of stage pairs rather than the full configuration space that end-to-end approaches must cover. 

\subsection{Deployment Optimization for AI Workflows}

Several systems formulate AI workflow deployment as constrained optimization. IPA optimizes variant selection, batch sizes, replicas, and resource allocation for linear inference pipelines as an integer program, using PAS as the accuracy objective~\cite{IPA}. Loki combines hardware and accuracy scaling for tree-structured pipelines using MILP-based resource allocation and runtime routing to reduce SLO violations~\cite{Ahmad2024Loki}. Both target homogeneous clusters and linear or tree-structured pipelines, modeling neither routed, feedback, nor composed compound AI topologies.
Other systems optimize deployment without treating accuracy as a joint objective. InferLine provisions and scales prediction pipelines under latency constraints with accuracy fixed externally~\cite{Crankshaw2020InferLine}. JellyBean deploys ML workflows across heterogeneous edge-to-cloud tiers, minimizing cost subject to throughput and accuracy SLOs~\cite{Wu2022JellyBean}.

Atlas differs along two dimensions. First, it formulates plan selection for compound AI workflows where variant selection, hardware placement, and topology jointly determine feasibility and predicted accuracy. Second, it uses MAP to estimate configuration-level accuracy from local conditional quality transitions, whereas existing systems either fix accuracy, compose standalone stage scores independently, or rely on end-to-end profiling.
\section{Discussion}
\label{sec:limitations}

Atlas provides an efficient middle ground between exhaustive profiling and product-based accuracy surrogates. However, it relies on several design assumptions. MAP models quality propagation through adjacent stage pairs as a first-order Markov process. This works when the upstream quality bucket captures the information most relevant to downstream behavior, as shown by the strong ranking results across the evaluated workflows. This assumption may be less accurate when later stages depend on outputs several hops earlier. Atlas supports sequential pipelines, routed workflows, loops, and their composition, covering common patterns such as RAG. Other workflow structures, such as fan-out aggregation workflows, would require additional operators. MAP also assumes that each stage exposes an intermediate quality signal that can be discretized into buckets. For stages without such a signal, task-specific instrumentation may be needed before transitions can be profiled. Finally, MAP reduces profiling cost substantially compared to exhaustive profiling, but it does not eliminate profiling entirely. For small configuration spaces, exhaustive profiling may remain simpler, while Atlas pays off for larger workflows where exhaustive profiling becomes impractical and product surrogates fail to preserve configuration rankings.


\section{Conclusion}
\label{sec:conclusion}

Deploying Compound AI workflows requires selecting an execution plan that maximizes accuracy under SLO constraints. Accurate estimation of workflow accuracy is the central obstacle in Compound AI deployment optimization. Atlas addresses this by separating accuracy estimation from system profiling and keeping both tractable. MAP profiles conditional quality transitions between adjacent stages, discretizes intermediate outputs into quality buckets, and composes the transitions according to the workflow topology. The Atlas optimizer selects execution plans with a MILP over the predicted accuracy and the SLO constraints. Across four Compound AI workflows, MAP achieves Spearman correlation up to $0.947$ and reduces profiling cost by 2.6$\times$ relative to exhaustive profiling. Critically, MAP's ranking advantage over product based approaches grows with workflow complexity, where deployment optimization matters most. Guided by MAP, the Atlas optimizer selects execution plans within 0.03 of oracle accuracy across all evaluated SLOs, while heterogeneous tier placement matches oracle accuracy at up to 42\% lower deployment cost than homogeneous placement strategies.

Future work will extend MAP with operators supporting various topologies, such as fan-out. Additionally, MAP accuracy modeling will be explored to support dynamic adaptation mechanisms, such as runtime model selection under varying workloads. Finally, Atlas will incorporate dynamic re-optimization, enabling execution plans to be revised as cluster conditions change without requiring full re-profiling.

\section*{Acknowledgment}

This work was partly funded by the European Union under the Horizon Europe programme through the SNS JU (Grant Agreement No. 101192912, NexaSphere). Views expressed are those of the authors and do not necessarily reflect those of the EU or the SNS JU.

\bibliographystyle{IEEEtran}
\bibliography{references}

\balance

\end{document}